\documentclass {article}

\usepackage{amssymb}
\usepackage{graphicx}

\usepackage{url}

\begin{document}


\title{Photonic Communications and Information
Encoding in Biological Systems }


%
%
\author{S.Mayburov\\
Lebedev Institute of Physics,\\
Leninsky pr. 53, Moscow, Russia\\
 E-mail: mayburov@sci.lebedev.ru}
\date{}
 \maketitle

\begin{abstract}

The structure of optical radiation emitted by the samples of loach
fish eggs is studied. It was found earlier that
 such radiation perform the communications between distant samples,
 which result in the synchronization of their development.
   The photon radiation in form of short quasi-periodic bursts was observed for fish and frog
   eggs, hence the communication mechanism can be  similar  to the exchange
 of binary encoded data in the computer nets via the noisy channels.
 The data analysis of fish egg radiation
  demonstrates that in this case the  information  encoding
is similar to the digit to time analogue algorithm.\\
\\

\end{abstract}


\section  {Introduction}

 Currently, the term 'biophotons' is attributed  to the optical and UV photons emitted by the living bio-systems
 in the processes which are different  from standard chemi-luminescence.
 Their systematic measurements  by the low-noise electronic photo-detectors was started about 1978  \cite {1,2};
 the  biophoton production (BP)  in optical and close UV range
 was established now for  large amount of  bio-systems \cite {3,B4}.
  It was  found that its rate and other parameters are quite sensitive to the characteristics of bio-system and its development.
  Because of it,  the biophoton measurements are applied now in many different fields from medical diagnostics
   to agriculture and ecology \cite {2}.

  The energy  spectrum  of biophotons is nearly constant  within optical and soft UV  range
   practically  for all studied bio-systems, so it  essentially differs from the  spectra expected
 for the system with the  temperature about $300^0$ K,
 which in this range should fall on 15 orders of magnitude \cite {2,3}.
The detailed BP mechanism is still unknown, but such excitations
can be stipulated by the biochemical reactions, in which oxygen
atoms are bound to the proteins and acids \cite {2,3}.
         The typical bio-photon rates are quite low, however,
the multiple experiments evidence that such radiation can perform
the effective signaling between distant bio-systems.  In
particular, being radiated by the growing organism or plant
 and absorbed by the similar one at the distance about several cm, it can rise the rate
 of cell division (mitosis) in it up to $30\%$ relative to the standard values. This phenomenon called mitogenetic effect (ME)
   is extensively studied in the last years  \cite {2,3}. Note that the  artificial constant illumination
    by the  visible light, even $10^4$ times more intense, can't induce the comparable gain.
The communications of some other types were reported also; for
the bio-systems in the state of abrupt stress or slow destruction
(apoptosis)  such radiation can change the state of other
bio-systems in the similar depressive way \cite {2,D4}.

Until now, ME and other  biophoton properties can't be described
within the standard framework of cellular biology.
 In our previous paper the  model of information exchange between the bio-systems
 by means of optical radiation was proposed  \cite {C4}. We've assumed that the main features
  of  such  communications can be similar to the  information exchange between the distant computers
   by the binary encoded messages. This hypothesis is prompted by the
   experiments
    which show that the radiation of some species  consists of  narrow quasi-periodic bursts
      (fig. 1), so its time structure is
  similar to the sequence of  electronic or photonic  pulses which transfer
  information in the computer communication channels \cite {1,B4}. To check our model in more detail
  we performed the
  data analysis for the radiation from loach fish eggs measured
   by the photomultipliers \cite {1,3}. Here some our results on the radiation structure are described,
   in particular,  the possible algorithm of photon signal encoding in the fish egg communications
   is obtained.

\section { Model of Bio-System Communications}

Before considering the photon exchange between the distant
bio-systems, it's worth to discuss how such communications can be
released inside the same dense bio-system. The optical and UV
excitations in the dense media  exist  as the quasi-particles
called excitons which can spread freely through the whole media
volume \cite {5}. They are strongly coupled  with electromagnetic
field, so they can be effectively produced during the photon
absorption by the media, the inverse process results in the
photons emission from  the system volume. It's established
experimentally now that the excitons play the important role in
the energy transfer inside the bio-systems, in particular, during
the photosynthesis  in plants and bacteria \cite {4,6}. Photon
production related to the nonlinear excitons in  protein
molecules was studied in \cite {7}. In our model the excitations
of biological media as the whole play the main role in biophotons
generation and absorption by the bio-system. We don't considered
 any model dynamics,
however, for such distances it's inevitably should have the
solitonic properties \cite {5}.
Hence our model supposes that the excitons spread freely over all
the volume of bio-system. In that case the exciton exchange can
constitute the effective system of signaling and regulation of
the bio-system development. The experiments evidence  that such
long-distance signaling regulates effectively the plant growth,
preventing from the large fluctuations of its global form,
 i.e. defines their morphogenesis \cite {2,3}.

As was noticed above, BP rate is quite low, about $10
\,photons/cm^2sec$ from the surface of large, dense bio-system. If
the corresponding field isn't coherent, then it described as the
stochastic ensemble of  photons. Then at its best  the absorption
of single photon or narrow bunch of photons can be detected by the
bio-system as the single independent 'click' or one bit of
information, analogously to standard photodetector devices. This
is the photocounting regime of  electromagentic field detection
well-known in quantum optics \cite {8}. We suppose that the same
approach is applicable also for the excitons produced and
absorbed in the same bio-system. Under these assumptions, the
exciton signaling between two parts  of the same bio-system and
photon signalling between two distant bio-systems can be quite
similar.

\begin{figure}
\centering
\includegraphics[height=7.2cm]{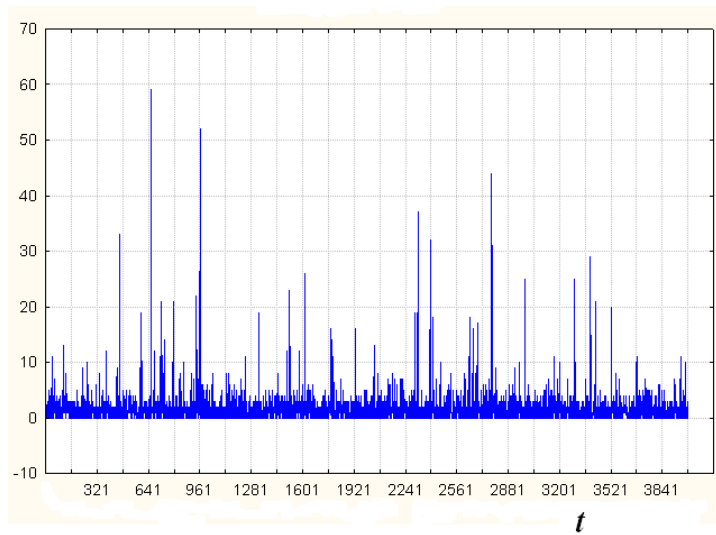}
\caption{Example of biophoton time spectra for development stage
16, full time scale 400 sec}
 \label{fig:example}
\end{figure}

It can be expected that the  signals which control the cell
mitosis and other functions  can be similar to the standard
discrete (binary) encoded messages transferred between two
computers  via the noisy communication channels \cite {C4}. The
origin of such similarity can be understood from the simple
reasoning without exploit of information theory machinery.
Plainly, for the  low exciton or photon radiation rates, typical
for bio-systems, the most important problem for the effective
signal transfer is to suppress the background. Even for the
bio-systems in complete darkness it induced by many sources,
like  the natural radioactivity, luminescent chemical impurities,
etc.. Consequently, as the main criteria characterizing  the
efficiency of information exchange, the signal to noise ratio
$K_O$ can be used, i.e. the ratio of registrated 'clicks' induced
by the bio-system signals and  the background.
 It's natural to suppose that the evolution of living
 species made the information exchange by means of
 photon radiation/absorption practically optimal.
 The
average rate of background radiation normally should be
 constant in time, so for given
bio-system with the limited  radiation intensity
 the optimal method  to achieve high $K_O$
level is to make the main bulk of the bio-system radiation to be
concentrated inside the short time intervals, i.e. the bio-system
radiation should  be in the form of bursts which would encode the
signals transferred to other bio-systems. The experiments with
fish eggs, fibroblast cells and other bio-systems demonstrated
that the biophotons are radiated by the short-time (less than 1
msec) quasi-periodic bursts \cite {1,B4}; the typical time
spectrum for fish eggs radiation is shown on fig. 1.

The influence of biophoton exchange on the organism development
was found for many species, in particular their detailed study was
performed for the eggs of loach fish  ( $Misgurnus
 fossilis$). Note that for the egg colony produced by the fish
during its breeding  the maximal rate of larvae survival is
achieved, if all eggs would develop with the same speed. However,
the small variations of temperature and water flow over colony
volume and other external factors tend to violate this condition.
It seems that the biophoton signaling between distant eggs  of
the same colony restore their simultaneous development. The
results for the optical contacts during 30-50 min between two
samples of fish eggs A,B of slightly different age demonstrate
the  significant synchronization of their development (\cite {1}
and refs. therein). However, it was found also that the optical
 contacts  between the fish eggs of significantly
  different ages result in  the serious violations of development in both
samples, for fish eggs at early stages the development can simply
stop. Those results evidence that the photon signals emitted by
fish eggs of different age can have the essentially different
structure, which encode the information about their age and
corresponding development program.

\section {Analysis of Experimental Data}

 The possible signal structure was exploited for the experimental
data on the optical radiation of loach fish eggs. The studied
sample is  the colony about 200 eggs,  which  is confined in the
quartz container filled with  water, their optical and soft UV
radiation from the container surface
 was registrated by the photo-multiplier.  Its
intensity was summed over the  nonoverlapping consequent time
intervals (bins) with the duration $\Delta=.1$ sec; the
experimental run normally consists of $6 *10^3$  such bins \cite
{1}. The measurements were performed for the  different
development stages, from the earliest ones (cleavage)  to the
latest 33-35 stages, preceding the free larvae appearance; the
average stage duration is about  1.5 hour. The background
radiation was measured for the empty container.

Since the background is supposedly stochastic and there is no time
correlations between its intensity at different time moments, then
the burst periodicity or some other time correlations between them
would help to discriminate the background more effectively.
Simultaneously, the variations of such correlations can also
encode the different signals send by the bio-system.
 It was supposed that such encoding is performed by
 the methods and algorithms
similar to the standard ones used for noisy communication
channels  \cite {C4}. In this framework, as the model example it
can be supposed that the separate message send by the bio-system
consists of $N$ bursts of the same height $I_r$ with time
interval $T_r$ between them, such messages are divided by the
periods of 'silence' $T_s$ when the detectable radiation is
similar to the stochastic background, i.e. each message
constitutes the burst cluster and such messages divided by $T_s$
intervals are repeated many times \cite {C4}. Plainly, for the
realistic bio-systems  those parameters would have some
stochastic spread $\sigma_I, \sigma_T,...$ around the average
values.

\begin{figure}
\centering
\includegraphics[height=7.2cm]{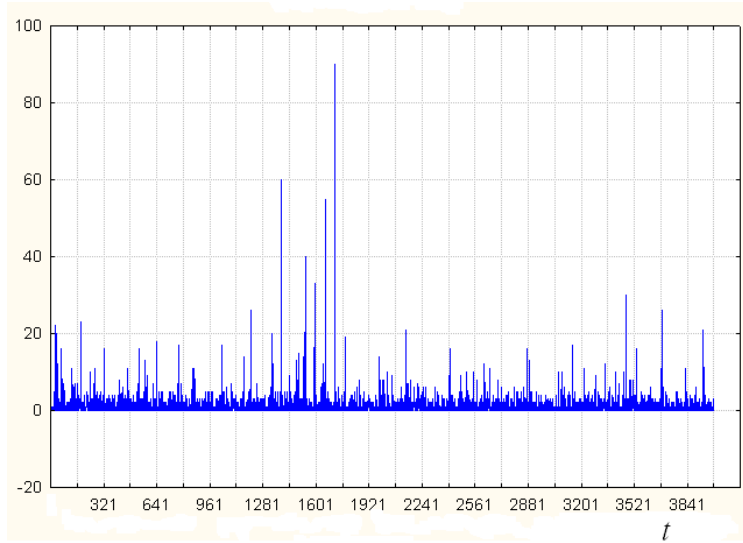}
\caption{Example of 'message' send by fish eggs at the development
stage 32, full time scale 400 sec}
 \label{fig:ex-2}
\end{figure}

First, we studied the discrimination of fish eggs radiation
against the background, for which the burst distribution should be
random with the arbitrary expectation values of burst amplitude
$\bar I$ and time interval $\bar T$ between the neighbor bursts.
The fish eggs radiation is expected to have more periodical
structure, for example,  if the burst radiation of fish eggs is
strictly periodical, i.e. it can be approximated as:
$$
    A(t)= I_r \delta(t-nT)
$$
 for arbitrary  integer $n$, then its fourier time spectra is equal
 to:
\begin {equation}
 a_{\nu}= \int\limits_0^{\infty} A(\tau)\cos \nu \tau \, d\tau = I_r
\delta(\nu-\frac{\pi l}{T})            \label {A}
\end {equation}
for integer $l$ and $0 < l \le \infty$. Thus,  the burst
periodicity is reflected in fourier spectra as the sequence of
periodical peaks, whereas for  background one can expect
$\bar{a}_{\nu}= \bar{I}$, i.e. is constant function with some
fluctuations around its average value. Hence Fourier analysis of
time spectra can be exploited to discriminate the periodic
radiation from the background; STATISTICA-7.0 program packet was
used for the data processing.

 The preliminary analysis of radiation data has shown that in terms of our model example the typical
duration of biophoton 'message' is about $10^2$ sec and $\bar {N}=
8.5$, i. e. $\bar T$ is about $12$ sec,  whereas $T_s$ is about
$4*10^2$ sec. From those estimates it follows that the density of
fourier time spectra
  $f(\nu)$ should be calculated for each consequent set of $1200$ bins in given run separately,
  i.e. $5$ spectra for each run. In this template the density
  $f(\nu)=a^2_{\nu}+b^2_{\nu}$ where $b_{\nu}$ is the value of
  corresponding  fourier integral (\ref {A})
  for $\sin \nu \tau$.
  Then for each spectra $f$ its autocorrelation:
  \begin {equation}
 g_l=\frac{ \sum\limits_0^M f(i)f(i+l) }
  {  \sum\limits_0^{M} f^2(i)}             \label {B}
\end {equation}
 was calculated for integer $i, l$, taking into account the finite length of run and
 its discreteness $M=\pi \Delta^{-1}$.
The sum $R$ of $g_l$ modules over such set is used as the
selection criteria:
$$
  R= \sum\limits_0^M |g_l|
$$
It was found that for arbitrary $R$ threshold which selects about
$80 \%$ of fish eggs runs, only $16 \pm 4 \%$ backgorund runs are
selected for $12 - 16$ stages ($10 \div 15$ hours after
fertilization) and $23 \pm 6 \%$ for $30 \div 34$ stages. It
evidences that the fish eggs radiation has more periodical
structure than the stochastic background, the possible candidate
for such periodical signal (message) is shown on fig. 2.
After such selection it was found that
 the individual messages have the length about $.6
\div 1.5*10^2$ sec and they are interspersed by the periods of
silence about $3 \div 6*10^2$ sec. Each message consists of $6 -
14$ distinct high bursts, despite significant fluctuations, their
amplitudes $I$ demonstrate gaussian-like dependence
 on the time distance from the message centre (see fig. 2).
 In each individual message the time intervals $T$ between the neighbor bursts with $I$ higher
than some threshold $I_0$ are nearly the same with some
dispersion or differ by the whole number, i.e.  are equal to $2T,
3T$, etc.. Yet $T$ average value can differ from one message to
another significantly, about factor $1.6$.

Concerning the dependence of signal form  on development stage,
our analysis  indicates that the most pronounced change suffers
$\bar T$  value. To demonstrate it, the inclusive (i. e. over all
run)
 $T$ distributions for the bursts with $I$ higher than $I_0=15$ units (see fig. 1)
were obtained for several development stages. To enlarge
statistics, we summed in the same plot the data for $8 - 11$, $12
- 16$ (fig. 3) and $30 - 34$ stages (fig. 4). Under these
conditions the total exposition time for each plot was 130 min..
For each distribution the systematic errors are negligible, the
statistical errors for each experimental point are equal to
$N^{\frac{1}{2}}_{events}$, where $N_{events}$ is the abcyssa
value in this point.  The obtained $\bar T$ expectation values
are equal to $\bar{T}=11.2 \pm .3$, $\bar{T}=9.2 \pm .3$ and $7.1
\pm .2$, sec for $ 8 \div 11$, $12 \div 16$ and $30 \div 34$
stages correspondingly.
   Meanwhile, the average
number of bursts is nearly the same for all three cases, whereas
the average burst amplitude $\bar I$ for $30 - 34$ stages is about
 $15 - 20\%$ larger than for $8 - 11$ stages, yet the overlap of
 their $I$ distributions is essentially larger than for $T$ distributions.

 The essential feature of all three plots is the presence of
 several large peaks, which are effectively higher than
 the possible statistical error limits. Their parameters
 need further analysis, but it's worth to notice that
 for $8 - 11 $ and $12 - 16$ stages two most prominent peaks
 seems to be generically connected, shifting from $5$ to $6$ sec
 and from $10.3$ to $10.8$ sec. Note also that $T$ maxima positions
 of three largest peaks for $8 -11$ stages are related
 approximately as $1:2:4$.

 The obtained distinctions explain, probably, how
the radiation from fish eggs of different age can influence the
'detector' samples in a different way. Despite that the
difference of signal parameters is only statistical, the multiple
repetition of such messages can be eventually percepted  by the
'detector' fish eggs as the different instructions which would be
fulfilled during their subsequent development. It seems that the
most important for their encoding is $T$ difference, because the
burst amplitude produce mainly the threshold effect, i.e. only
the condition like $I > I_0$  is accounted for some arbitrary
$I_0$ which is proper for given bio-system. If those conclusions
will be confirmed by further experiments, it would mean that the
main encoding algorithm for fish eggs radiation has the analogue
realization, despite the produced signal is constituted by the
sequence of discrete bursts. Such information encoding is similar
to the digital-time analogue (DTA) algorithm used in some
electronic systems. Note that the similar encoding of electric
pulses supposedly is exploited in the  brain neuron chains \cite
{6}.

\begin{figure}
\centering
\includegraphics[height=6.6cm]{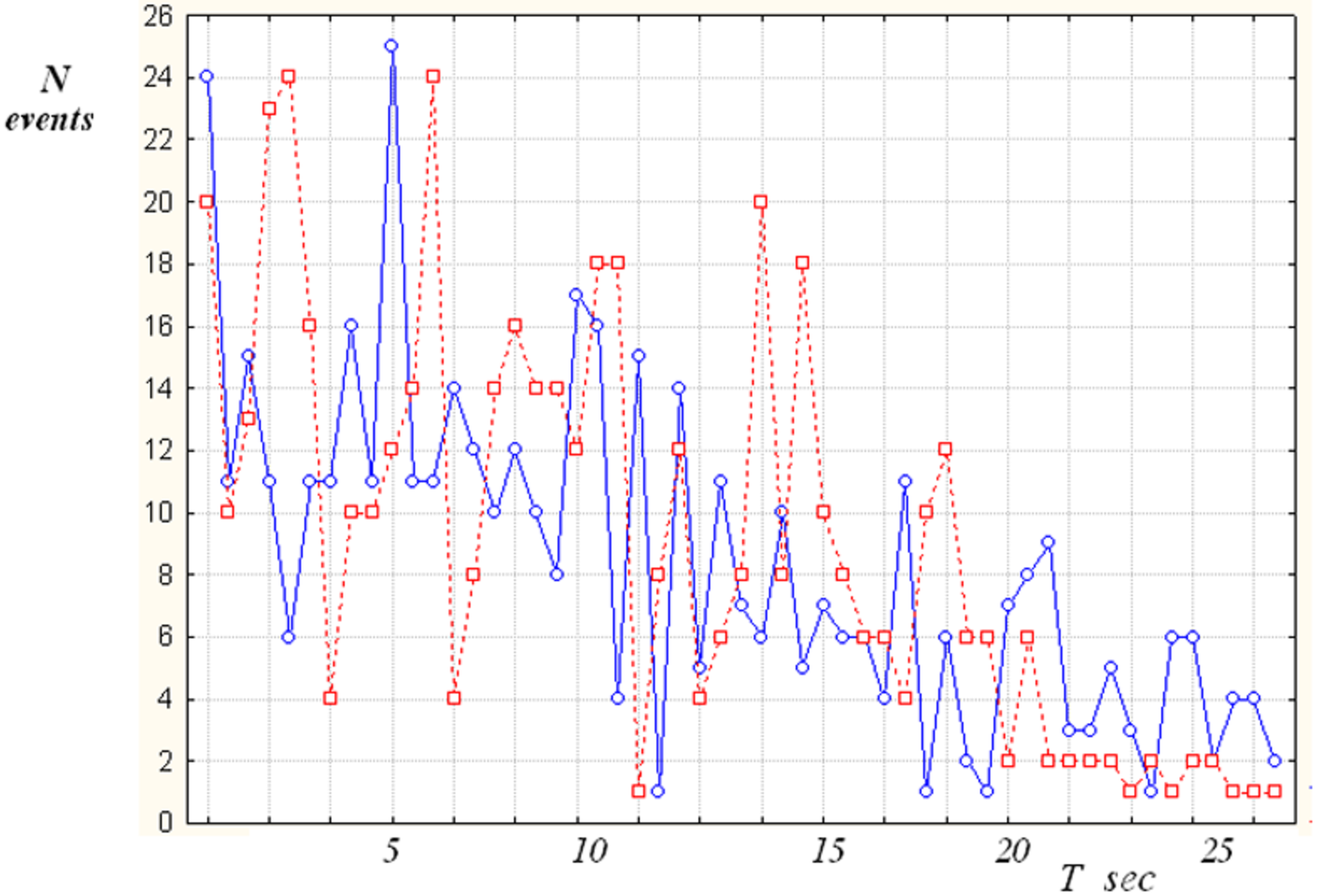}
\caption{$T$ distribution for fish eggs radiation; solid line:
 $8 \div 11$  stages; broken line  $12 \div 16$ stages}
 \label{fig:ex-3}
\end{figure}

\begin{figure}
\centering
\includegraphics[height=7.5cm]{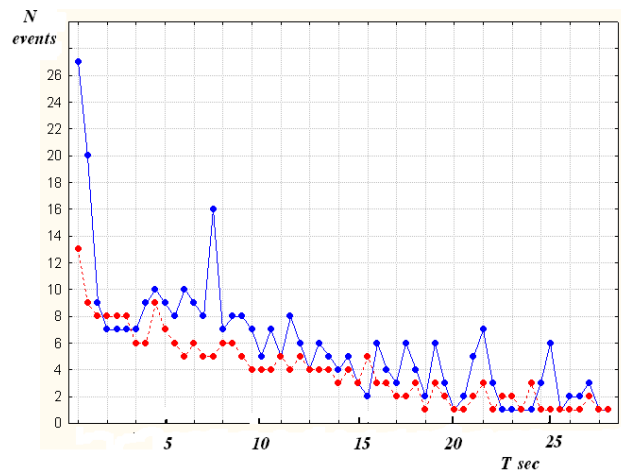}
\caption{$T$ distribution for fish eggs radiation; solid line:
 $30 \div 34$  stages; broken line - background}
 \label{fig:ex-4}
\end{figure}

%


Our model assumes that the bio-system's field is noncoherent, yet
it's worth to consider also  the possibility that this e-m field
of bio-system
          can possess the spacious short-time coherence within the observed photon bursts,
           similarly  to  the coherence of laser pulse.
           Some experiments evidence  that such biophoton coherence really takes place \cite
           {9}; in this experimental  set-up the transparent quartz plate was
             installed
             between the inductor and detector bio-systems. In the first run
              the plate parallel surfaces were smooth and polished, so that it doesn't
              perturb the phase relations between the different pieces of wave front
               of incoming photons. In another run, the plate has the random deflections
              from the surface parallelism, which violated such phase relations,
               and so  resuly in the violation of the impact  wave coherence. It was found that in comparison
                with the control sample of isolated bio-system, the radiation passed
                 through the random surface results in the gain of mitosis rate of $20 \%$,
                   yet for smooth surface it reach the rate of $45 \%$.
               In our framework, it's reasonable to suppose that the field
                with such 'transversal' coherence more effectively produce
                the collective excitations in the cells cluster, than noncoherent field.
                  If this hypothesis is correct, then such
                  coherence effects
                   will not change the principal scheme of communications proposed here,
                    rather, it would enlarge its efficiency.

 For the conclusion the obtained results can help to reveal the mechanism of
communications between the distant samples of fish eggs and permit
to describe the universal features of biophoton signaling between
the separate bio-systems. The similar mechanism can, on the all
appearances, describe the exciton signaling inside the  dense
bio-system. The  cell signaling  and regulation features are well
studied for the extracellular  biochemical reactions \cite {6}.
Concerning the chemical signaling in the tissues, its efficiency
 and precision is principally restricted by the molecular
 diffusion effects inside the bio-system media and so
 can transfer the signals only for small distances.
 Note also that
 the exciton signaling inside organism can be much faster, than
 the  chemical one by means of molecular messangers.
 Hence it can be efficient in case
 of stress or the abrupt change of external conditions.
 Experimental results show that under the different stress
 conditions the photon rates from bio-system can rise in short
 time significantly, probably, as the consequence of intensive
 internal signaling  \cite {2,3}.

 Author is thankful L.V. Beloussoff and I.V. Volodiaev for
 providing the experimental data files and the extensive consultations on
 the related topics.

\newpage

\end {document}